\documentclass[preprint,aps,showpacs,nofootinbib,preprintnumbers,amsmath,amssymb]{revtex4-1}
\usepackage{epsfig}

\usepackage{dcolumn}
\usepackage{bm}
\usepackage[usenames ,dvipsnames]{xcolor}
\usepackage{slashed}
\usepackage{float}
\usepackage{graphicx,color}
\usepackage{amsmath}
\usepackage{nonfloat}

\begin{document}
	
\title{ Charmed Baryon Weak Decays with Vector Mesons}

\author{C.Q. Geng$^{1,2,3,4}$, Chia-Wei Liu$^{3}$ and Tien-Hsueh Tsai$^{3}$}
\affiliation{
$^{1}$School of Fundamental Physics and Mathematical Sciences, Hangzhou Institute for Advanced Study, UCAS, Hangzhou 310024, China \\
$^{2}$International Centre for Theoretical Physics Asia-Pacific, Hangzhou 310024, China \\
$^{3}$Department of Physics, National Tsing Hua University, Hsinchu 300, Taiwan\\
$^{4}$Physics Division, National Center for Theoretical Sciences, Hsinchu 300, Taiwan
}\date{\today}

\begin{abstract}
We give a systematic study of ${\bf B}_c\to {\bf B}_n V$  decays, where ${\bf B}_c$  and $ {\bf B}_n$ correspond to the anti-triplet charmed and octet baryons,  respectively, while $V$ stand for the vector mesons. We calculate the  color-symmetric contributions to the decays from the effective Hamiltonian with the factorization approach and extract the anti-symmetric ones based on the experimental measurements and  $SU(3)_F$ flavor symmetry. We find  that most of the existing experimental data for ${\bf B}_c\to {\bf B}_n V$ are consistent with our fitting results. We present all the  branching ratios of the Cabbibo allowed, singly Cabbibo suppressed and doubly Cabbibo suppressed decays of  ${\bf B}_c\to {\bf B}_n V$. The decay parameters for the daughter baryons and mesons in ${\bf B}_c\to {\bf B}_n V$ are also evaluated. In particular, we point out that the Cabbibo allowed decays of $\Lambda_c^+ \to \Lambda^0 \rho^+$ and $ \Xi_c^0 \to \Xi^- \rho^+$ as well as the singly Cabbibo  suppressed ones of  $\Lambda_c^+ \to \Lambda^0 K^{*+}$, $\Xi_c^+ \to \Sigma^+ \phi$ and $\Xi_c^0\to \Xi^- K^{*+}$
have large branching ratios and decay parameters with small uncertainties, which can be tested by the  experimental searches at the charm facilities.
\end{abstract}
\maketitle

\section{Introduction}
Recently, the LHCb Collaboration has obtained the anti-triplet charmed baryon lifetimes with high precision, given by~\cite{Lifetime}
\begin{equation}\label{times}
(\tau_{\Lambda_c^+}, \tau_{\Xi_c^+}, \tau_{\Xi_c^0} )= (203.5\pm 2.2~ , 456.8\pm 5.5~, 154.5\pm 2.5 )~\text{fs}\,.
\end{equation}
Note that the decay lifetime of $\Xi_c^0$ is 3$\sigma$ above the previous averaged value of
$(112\pm12)$ fs in PDG~\cite{pdg}. 
Furthermore, 
BES\MakeUppercase{\romannumeral 3}~\cite{Ablikim:2015flg} and Belle~\cite{Zupanc:2013iki} 
Collaborations have precisely measured 
 the absolute decay branching ratio for $\Lambda_c^+ \to pK^-\pi^+$ with the world average of
\begin{equation}
{\cal B}(\Lambda_c^+ \to p K^- \pi^+) = (6.28\pm 0.32) \%
\end{equation}
in PDG~\cite{pdg}. Moreover, the Belle Collaboration has determined the absolute branching ratios in $\Xi_c$, 
given by~\cite{Exp:absXic0,Exp:absXicp}
\begin{eqnarray}
{\cal B}(\Xi_c^0\to \Xi^- \pi^+) &=& (1.80\pm 0.52)\%\,,\nonumber\\
{\cal B}(\Xi_c^+ \to \Xi^- \pi^+\pi^+) &=& (2.86\pm 1.27) \% \,,
\end{eqnarray}
from the decay chains of $B$ mesons. These decay branching ratios are important as  most of the other branching ratios of anti-triplet charmed baryons are measured relative to them.

It is known that there are some difficulties for the  theoretical study in the  non-leptonic decays of charmed baryons
due to the failure of  the  factorization approach. On the other hand, one can use the $SU(3)_F$ flavor symmetry
to relate the amplitudes  among different decays~\cite{Sharma:1996sc,Savage:1989qr,Savage:1991wu}. 
This becomes possible~\cite{Roy:2019cky,Lu:2016ogy,first,zero,second,third,fourth,fifth,Wang:2017gxe,sixth,Zhao:2018mov,Hsiao:2019yur,Geng:2019bfz,Latest_three,SU(3)Dec,He:2018joe,Jia:2019zxi,Chen:quark}
 as there have been recently many new 
experimental measurements for charmed baryon decays~\cite{Zupanc:2013iki,Exp:absXic0,Exp:absXicp,Ablikim:2015flg,Ablikim:2015prg,Ablikim:2016tze,Ablikim:2016mcr,Ablikim:2017ors,
Ablikim:2016vqd,Ablikim:2017iqd,Ablikim:2018jfs,Ablikim:2018bir,etaAb,SiAbsoluteBr}. 
In addition to the  analysis of charmed baryon decays with $SU(3)_F$, 
the theoretical calculations based on dynamical models have also been done in the literature~\cite{Xu92,xialpha1,Zenczykowski:1993jm,Cheng:allow,Cheng:general,Uppal:1994pt,Verma98,pole,Chen:2002jr,Cheng:sup,Korner:1992wi,Cheng:latest,Wang:2017mqp}.
However, the results are often not reliable and different among models. 
The main difficulties are due to the unknown baryon wave functions and nonfactorizable contributions.

In this work, we concentrate on the decays of ${\bf B}_c\to {\bf B}_n V$ with the $SU(3)_F$ flavor symmetry, 
where ${\bf B}_c$ and ${\bf B}_n$ correspond to the anti-triplet charmed and octet baryons,  and $V$ stand for the vector mesons, respectively.
In fact,  some of the decay branching ratios have been recently explored  based on $SU(3)_F$ in Ref.~\cite{Hsiao:2019yur}.
However, the approach in Ref.~\cite{Hsiao:2019yur} has ignored  the contributions from color-symmetric parts of 
the effective Hamiltonian  and  correlations among the $SU(3)_F$ parameters. 
In addition, there should be four independent wave amplitudes~\cite{SPD}, but only one is used in Ref.~\cite{Hsiao:2019yur}. 
In this study, we shall include all the wave amplitudes and consider the full effective Hamiltonian.
We shall also discuss  the decay asymmetry parameters in ${\bf B}_c\to {\bf B}_n V$, such as the up-down and longitudinal polarization
asymmetries of ${\bf B}_n $ and asymmetry parameter of $V$.


This paper is organized as follow. In Sec.~\MakeUppercase{\romannumeral 2}, we present the formalism.
In Sec.~\MakeUppercase{\romannumeral 3}, we extract the $SU(3)_F$ parameters from the experimental data.  We conclude our study in Sec~\MakeUppercase{\romannumeral 4}.

\section{Formalism}
The most general form of the amplitude for ${\bf B}_c \to {\bf B}_n V$ can be written as 
\begin{equation}\label{general_amp}
M=\bar{u}_{f}\left(p_{f}\right) \epsilon^{ \mu*}\left[A_{1} \gamma_{\mu} \gamma_{5}+A_{2}\frac{p_{f \mu}}{m_i} \gamma_{5}+B_{1} \gamma_{\mu}+B_{2} \frac{p_{f \mu}}{m_i}\right] u_{i}\left(p_{i}\right)\,,
\end{equation}
where $\epsilon^\mu$ is the four vector polarization for the vector meson of  $V$, $u_i(p_i)$ and $u_f(p_f)$ are the 4-component spinors (momenta) for 
the initial and final baryons, respectively, and $m_i$ represents the initial baryon mass.
In general, the physical vector meson with its momentum in the $z$ direction has the vector polarizations of
 $\epsilon^\mu=(0,\frac{1}{\sqrt{2}},\frac{\pm i}{\sqrt{2}},0)$ for $\lambda_V=\pm 1 $ and $\epsilon^\mu = (|\vec{p}_V|/m_V,0,0,E_V/m_V)$ for $\lambda_V = 0$, where $\lambda_V$ is the helicity and $m_V$, $\vec{p}_V$ and $E_V$ are the mass, 3-momentum and energy of the vector meson, respectively.
In the center of the momentum frame~(CMF), the kinematic factors of $A_2$ and $B_2$ in Eq.~(\ref{general_amp}) can be further written as 
\begin{equation}\label{A2supp}
\epsilon^{\mu*} p_{f\mu}/m_i = \epsilon^{ \mu*} p_{i\mu}/m_i =\epsilon^{0*} \,.
\end{equation}
Here, we have used $p^{\mu}_i =p^{\mu}_f+p^{\mu}_V$ and $\epsilon_\mu p_V^{\mu} = 0$, where $p_V$ corresponds to the 4-momentum of the
vector meson. 
It is clear that the terms associated with $A_2$ and $B_2$ will only contribute to the decay in the case of $\lambda_V=0$, which are suppressed by the factor of $p_c/m_V$ with $p_c$ defined as the magnitude of the 3-momentum in the CMF, so that they can be ignored.	

The decay width, up-down asymmetry  and longitudinal polarization of 
${\bf B}_c \to {\bf B}_n V$
are given by
\begin{eqnarray}
\Gamma&=&\frac{p_{c}}{4 \pi} \frac{E_{f}+m_{f}}{m_{i}}\left[2\left(|S|^{2}+\left|P_{2}\right|^{2}\right)+\frac{E_{V}^{2}}{m_{V}^{2}}\left(|S+D|^{2}+\left|P_{1}\right|^{2}\right)\right]\,,\\
\alpha&=&\frac{2 E_{V}^{2} \operatorname{Re}(S+D)^{*} P_{1}+4 m_{v}^{2} \operatorname{Re}\left(S^{*} P_{2}\right)}{2 m_{V}^{2}\left(|S|^{2}+\left|P_{2}\right|^{2}\right)+E_{V}^{2}\left(|S+D|^{2}+\left|P_{1}\right|^{2}\right)}\,,\\
P_L&=&\frac{2 E_{V}^{2} \operatorname{Re}(S+D)^{*} P_{1}-4 m_{V}^{2} \operatorname{Re}\left(S^{*} P_{2}\right)}{2 m_{V}^{2}\left(|S|^{2}+\left|P_{2}\right|^{2}\right)+E_{V}^{2}\left(|S+D|^{2}+\left|P_{1}\right|^{2}\right)}\,,
\end{eqnarray}
where 
$S$, $P_{1,2}$ and $D$, corresponding to the orbital angular momenta of $l=0,1,2$ in the non-relativistic limit, 
are given by~\cite{SPD}
\begin{eqnarray}
S&=&-A_{1}\,,\\
P_{1}&=&-\frac{p_{c}}{E_{V}}\left(\frac{m_{i}+m_{p}}{E_{f}+m_{f}} B_{1}+ B_{2}\right)\,,\\
P_{2}&=&\frac{p_{c}}{E_{f}+m_{f}} B_{1}\,,\\
D&=&-\frac{p_c^2}{E_V(E_f+m_f)}\left(A_{1}- A_{2}\right)\,,
\end{eqnarray}
respectively.
Here, $\alpha$ and $P_L$ are defined by
\begin{eqnarray}
\frac{d\Gamma}{d \cos \theta} &\propto& 1 + \alpha \cos \theta \,,\\
P_L &=&\frac{\Gamma(\lambda_{{\bf B}_n}=1)-\Gamma(\lambda_{{\bf B}_n}=-1)}{\Gamma(\lambda_{{\bf B}_n}=1)+\Gamma(\lambda_{{\bf B}_n}=-1)} \,,
\end{eqnarray}
where $d\Gamma$ is the partial decay width, $\lambda_{{\bf B}_n}$ is the helicity of ${\bf B}_n$ and $\theta$ is the angle between the
spin and momentum directions of ${\bf B}_c$ and ${\bf B}_n$, respectively.

Since the vector meson of $V$ subsequently decays into two pseudo-scalar mesons, its polarization can be determined. As a result, we can discuss the decay asymmetry parameter of $V$, defined by~\cite{Korner:1992wi}
\begin{equation}
\frac{d\Gamma_V}{d\cos \theta_V}\propto 1+\alpha_V \cos ^2 \theta_V\,,
\end{equation}
with
\begin{equation}
\alpha_V=\frac{E_{V}^{2}\left(|S+D|^{2}+\left|P_{1}\right|^{2}\right)- m_{V}^{2}\left(|S|^2+|P_2|^2\right)}{ m_{V}^{2}\left(|S|^2+|P_2|^2\right)}\,,
\end{equation} 
where $d\Gamma_V$ is the partial decay width for the $V$  decay and $\theta_V$ is the polar angle between $\vec{p}_V$ and the momentum directions of the pseudo-scalar mesons in the CMF of $V$.

The effective Hamiltonian responsible  for the decay processes with $\Delta c = -1$ is
\begin{equation}\label{effectHa}
{\cal H}_{eff} = \frac{G_F}{\sqrt{2}}\sum_{q,q^\prime=d,s}\left[ c_1V_{uq}^*V_{cq^\prime}(\bar{u}q)(\bar{q^\prime}c) + c_2V_{uq}^*V_{cq^\prime}(\bar{u}q)(\bar{q^\prime}c) \right]\,,
\end{equation}
where the quark operators are defined as $(\bar{q}_1 q_2)=(\bar{q}_1\gamma_\mu(1-\gamma_5) q_2)$ with summing over the colors, 
the Wilson coefficient of  $c_1(c_2)$ is $1.246(-0.636)$ at the scale of $\mu = 1.25$~GeV~\cite{Buras:1998raa} and  $G_F$ is the Fermi constant. 
Note that $(q,q')=(d,s)$, $(d, d)$ or $(s,s)$ and $(s,d)$ correspond to the Cabbibo allowed, singly Cabbibo suppressed and doubly Cabbibo
suppressed  decays, respectively.

By using the CKM mixing parameters of $V_{cs}=V_{ud}\approx 1$ and  $s_c\equiv V_{us} = -V_{cd}\approx 0.225$,  the effective Hamiltonian in the flavor basis
is given by
\begin{equation}
{\cal H}_{eff} = \frac{G_F}{2\sqrt{2}}\left( \frac{c_-}{2}\epsilon^{lij}H(6)_{lk} + c_+ H(\overline{15})^{ij}_k \right)(\bar{q_j}q^k)(\bar{q_i} c)
\end{equation}
where $(q_1,q_2,q_3) = (u,d,s)$, $c_\pm=c_1\pm c_2$,
and  $\epsilon^{lij}$  represents the total antisymmetric tensor with $\epsilon^{123} =1$.
Here, the tensor components are given by
\begin{eqnarray}
H(6)_{ij}&=&\left(
\begin{array}{ccc}
0& 0 & 0\\
0 & 2 & 2s_c\\
0 & 2s_c& 2s_c^2
\end{array}\right)\,,\nonumber\\
H(\overline{15})^{ij}_k&=&
\left(\begin{array}{ccc}
\left(\begin{array}{ccc}
0&0&0\\
0&0&0\\
0&0&0
\end{array}\right),
\left(\begin{array}{ccc}
0&-s_c&1\\
-s_c&0&0\\
1&0&0
\end{array}\right),
\left(\begin{array}{ccc}
0&-s_c^2&s_c\\
-s_c^2&0&0\\
s_c&0&0
\end{array}\right)
\end{array}\right)\,.
\end{eqnarray}
Two of the creation operators generated by $H(\overline{15})$ are symmetric in color. 
As a result,  $H(\overline{15})$ does not contribute to the nonfactorizable amplitudes since the charmed baryons 
are total anti-symmetric in color~\cite{Korner,Pati:1970fg}.

We separate  $A_1$ and $B_1$ into $6$ and $\overline{15}$ parts under the $SU(3)_F$ symmetry:
\begin{eqnarray}
\label{eq18}
A_1= A_1^{(6)}+A_1^{(\overline{15})}\,,\nonumber\\
\label{distin}B_1= B_1^{(6)}+B_1^{(\overline{15})}\,.
\end{eqnarray}
In Eq.~(\ref{eq18}), $A_1^{(6)}$ and $B_1^{(6)}$ are parametrized as 
\begin{eqnarray}\label{A_1(6)}
A_1^{(6)}({\bf B}_c\to{\bf B}_n V) =&& a_0 H(6)_{ij}({\bf B}_c')^{ik}({\bf B}_n)^j_k (V)^l_l 
+a_1 H_{ij}(6)({\bf B}_c')^{ik}({\bf B}_n)_k^l (V)_l^j \nonumber\\
&&+a_2 H_{ij}(6)({\bf B}_c')^{ik}(V)_k^l({\bf B}_n)_l^j+\label{A2}a_3 H_{ij}(6)({\bf B}_n)_k^i (V)_l^j ({\bf B}_c')^{kl}\,, \\
B_1^{(6)}({\bf B}_c\to{\bf B}_n V) =&& b_0 H(6)_{ij}({\bf B}_c')^{ik}({\bf B}_n)^j_k (V)^l_l 
+b_1 H_{ij}(6)({\bf B}_c')^{ik}({\bf B}_n)_k^l (V)_l^j \nonumber\\
&&+b_2 H_{ij}(6)({\bf B}_c')^{ik}(V)_k^l({\bf B}_n)_l^j
\label{B22}+b_3 H_{ij}(6)({\bf B}_n)_k^i (V)_l^j ({\bf B}_c')^{kl}\,,
\end{eqnarray}
where $a_i$ and $b_i$ are the $SU(3)_F$ parameters, while ${\bf B}_{c,n}$ and $V$ can be written under
the tensor components of the $SU(3)_F$ representations, given by
\begin{eqnarray}
&&{\bf B}_c = (\Xi_c^0 , - \Xi_c^+ , \Lambda_c^+)\nonumber\\
&& {\bf B}_n=\left(\begin{array}{ccc}
\frac{1}{\sqrt{6}}\Lambda+\frac{1}{\sqrt{2}}\Sigma^0 & \Sigma^+ & p\\
\Sigma^- &\frac{1}{\sqrt{6}}\Lambda -\frac{1}{\sqrt{2}}\Sigma^0  & n\\
\Xi^- & \Xi^0 &-\sqrt{\frac{2}{3}}\Lambda
\end{array}\right)\,,
\end{eqnarray}
and
\begin{eqnarray}
&&V=\left(\begin{array}{ccc}
\frac{1}{\sqrt{2}}(\omega+\rho^0 )  & \rho^+ & K^{*+}\\
\rho^- &\frac{1}{\sqrt{2}}(\omega-\rho^0) &  K^{*0}\\
K^{*-} & \bar K^{*0}&\phi
\end{array}\right)\,,
\end{eqnarray}
respectively.

On the other hand, the contribution from $H(\overline{15})$ to $c \to uq'\bar{q}$  is factorizable, given by
\begin{equation}
M(\overline{15})=\frac{G_{F}}{2\sqrt{2}}V_{uq}V_{cq'} c_+\left(1 +\frac{1}{N_c}  \right)\left\langle V|(\bar{u} q)| 0\right\rangle\left\langle{\bf B}_n|(\bar{q'} c)| {\bf B}_c\right\rangle
\end{equation}
for the vector mesons with positive charges, while the creation operators, $\bar{q'}$ and $\bar{u}$, are interchanged for the neutral vector mesons. 
Accordingly, $A_1^{(\overline{15}) }$ and $B_1^{(\overline{15}) }$ in Eq.~\eqref{distin} are given by~\cite{Wang:2017mqp}
\begin{eqnarray}
\label{eq24}
A_{1}^{(\overline{15}) } &=&-\frac{G_F }{2\sqrt{2}}V_{uq}V_{cq'}f_Vm_Vc_+\left(1 +\frac{1}{N_c}  \right)\left( g_1 - g_2\frac{m_i-m_f}{m_i}\right)\,,\nonumber\\
B_{1}^{(\overline{15}) } &=&\frac{G_F }{2\sqrt{2}}V_{uq}V_{cq'}f_Vm_Vc_+\left( 1+\frac{1}{N_c}  \right)\left( f_1 + f_2\frac{m_i+m_f}{m_i}\right)\,,
\end{eqnarray}
where $\langle V | (\bar{q} q')| 0 \rangle = f_Vm_V\epsilon^*_\mu$  and 
 $N_c$ is the effective color number. 
 In Eq.~(\ref{eq24}),
 we take that $f_V = 0.215$~GeV and the form factors of $f_i(g_i)$ are defined by
 \begin{eqnarray}
\langle {\bf B}_n | (\bar{q}c)|{\bf B}_c\rangle &=& \bar{u}_f(p_f)\left[\left( f_1\gamma_\mu -f_2i\sigma_{\mu\nu}\frac{q^\nu}{m_i}+  f_3 \frac{q_\mu}{m_i}\right)\right.\nonumber\\
&&\left.- \left(  g_1 \gamma_\mu - g_2 i\sigma_{\mu\nu}\frac{q^\nu}{m_i}+g_3\frac{q_\mu}{m_i} \right)\gamma_5\right] u_i(p_i)\,.
 \end{eqnarray}
In our calculation, we evaluate the form factors from the MIT bag model~\cite{BAG,Mit}. 
The baryon wave functions and  form factors are listed in Appendix~A.

The factorizable parts in $A_2$ and $B_2$ are given by 
\begin{eqnarray}
A_2^{(fac)} &=& \frac{G_F}{\sqrt{2}} \left[ c_+\left(1 + \frac{1}{N_c}\right) \pm  c_-\left(1 - \frac{1}{N_c}\right)  \right]V_{uq}V_{cq'}f_V m_Vg_2\,,\\
B_2^{(fac)} &=&- \frac{G_F}{\sqrt{2}} \left[ c_+\left(1 + \frac{1}{N_c}\right) \pm  c_-\left(1 - \frac{1}{N_c}\right)  \right]V_{uq}V_{cq'}f_V m_Vf_2
\end{eqnarray}                         
with the ``$\pm$'' signs for mesons with positive and neutral charges, respectively.
In general, it is also possible to parametrize the nonfactorizable contributions in $A_2$ and $B_2$ according to the $SU(3)_F$ symmetry.
However, since they are suppressed  due to  Eq.~(\ref{A2supp}), we will neglect these parts. 

To sum up,  $A_1^{(\overline{15})}(B_1^{(\overline{15})})$ and $A_2^{(fac)}(B_2^{(fac)})$ can be  calculated from the factorization approach, while $A_1^{(6)}(B_1^{(6)})$ are parametrized by the $SU(3)_F$ symmetry. 
The detail results are shown  in Appendix~B.

\section{Numerical Results}
 The effective color number can be extracted from the decay
 branching ratio of $\Lambda_c^+ \to p \phi$ since it only receives the factorizable contribution~\cite{Cheng:sup,Korner}.
 The decay amplitude  is given by
\begin{equation}
M(\Lambda_c^+ \to p \phi)=\frac{G_{F}}{\sqrt{2}}V_{us}V_{cs} \left(c_2 +\frac{c_1}{N_c}  \right)\left\langle \phi |(\bar{s} s)| 0\right\rangle\left\langle p|(\bar{u} c)| \Lambda_c\right\rangle\,.
\end{equation}
With the form factors given in Appendix~A,
we obtain the decay parameters
\begin{equation}
\alpha (\Lambda_c^+ \to  p \phi) = -0.08\,\,,\,\,\,\,\,\,\,\,P_L (\Lambda_c^+ \to  p \phi) = -0.85\,\,,\,\,\,\,\,\,\,\,\alpha_V(\Lambda_c^+ \to  p \phi) =0.97\,, \
\end{equation}
which are independent of $N_c$. On the other hand, with the experimental data of  
${\cal B}(\Lambda_c^+ \to p \phi) = (1.06\pm 0.14)\times 10^{-3}$~\cite{pdg}, 
we find that $(c_2 +c_1/N_c )=0.49$,  leading to $N_c=9$, while the effective coupling strengths are found to be
\begin{eqnarray}\label{C_1}
A_1(\Lambda_c^+ \to p \phi)&=& 0.0110~G_F\text{GeV}^2\,,\,\,\,\,\,B_1(\Lambda_c^+ \to p \phi) = -0.0175~G_F\text{GeV}^2\,,\\
A_2 (\Lambda_c^+ \to p \phi)&=&0.0034~G_F\text{GeV}^2 \,,\,\,\,\,\,B_2 (\Lambda_c^+ \to p \phi)= 0.0109~G_F\text{GeV}^2\,.
\end{eqnarray}

For the other decay modes of ${\bf B}_c\to {B}_nV$, the nonfactorizable effects in $A_1^{(6)}$ and $B_1^{(6)}$ are sizable,
which cannot be ignored. The calculations of the nonfactorizable amplitudes are non-perturbative, which are generally model dependent. 
To tackle with these effects, we determine the parameters in Eqs.~(\ref{A2}) and (\ref{B22}) with the experimental data,
which  are listed in Table~\ref{Table1}. Here, we have used Eq.~(\ref{AbXip}) in Appendix~C to exact the branching ratios in $\Xi_c^+$.
In particular,  we have that ${\cal B}(\Xi_c^+ \to \Xi^0 \rho^+)\approx {\cal B}(\Xi_c^+ \to \Xi^0 \pi^+\pi^0)=(8.3\pm 3.6)\%$ as
the experimental branching ratio as stated in Appendix~C. 
In addition, the branching ratios of $\Xi_c^+ \to \Sigma^+ \phi$ and $ \Lambda_c^+\to \Sigma^+\rho^0$ can be obtained by the event counting method in  Refs.~\cite{SigmaPPhi,SigmaPrho0}.
For $\Lambda_c^+\to p \phi$,
we impose 10\% error deviations for  the effective coupling strengths in Eq.~(\ref{C_1})  to  account for the errors in the form factors evaluated from 
the MIT bag model.
\begin{table}
	\begin{center}
		\centering
		\caption{Decay branching ratios of ${\bf B}_c \to {\bf B}_n V$ from the
		experimental data and our $SU(3)_F$ reconstructed values.}\label{Table1}
		\begin{tabular}[t]{lll|lll}
			\hline
channel& $10^2{\cal B}_{ex}$&$10^2{\cal B}_{SU(3)}$& channel & $10^3{\cal B}_{ex}$&$10^3{\cal B}_{SU(3)}$\\
			\hline
$ \Lambda_{c}^{+} \to \Lambda^{0} \rho^{+} $&$ <6 $~\cite{pdg,LcLrhoold} &$4.81\pm 0.58$&$ \Lambda_{c}^{+} \to p \omega $&$ 0.94\pm 0.39 $~\cite{pdg}&$0.63\pm 0.34$\\
$ \Lambda_{c}^{+} \to \Sigma^{+} \omega $&$ 1.70 \pm 0.21 $~\cite{pdg}&$1.81\pm 0.19$&$ \Lambda_{c}^{+} \to \Sigma^{+} K^{*0} $&$ 3.5\pm 1.0  $~\cite{pdg}&$0.38\pm 0.09$\\
$ \Lambda_{c}^{+} \to p \bar{K}^{*0} $&$ 1.96 \pm 0.27 $~\cite{pdg}&$2.03\pm 0.25$&$ \Lambda_{c}^{+} \to p \phi $&$ 1.06\pm 0.14 $~\cite{pdg}&$0.87\pm 0.14$\\
$ \Lambda_{c}^{+} \to \Sigma^{+} \phi $&$ 0.39\pm 0.06 $~\cite{pdg}&$ 0.39\pm 0.06 $&$ \Xi_{c}^{+} \to p \bar{K}^{*0} $&$ 4.13\pm 1.69 $~\cite{pdg,Exp:absXicp}&$ 4.71\pm 1.22$\\
$ \Lambda_{c}^{+} \to \Sigma^{+} \rho^{0} $&$ 1.0 \pm 0.5 $~\cite{pdg}&$1.43\pm 0.42$&$ \Xi_{c}^{+} \to \Sigma^{+} \phi $&$ 1.17\pm0.87 $~\cite{pdg,SigmaPPhi,Exp:absXicp}&$1.82\pm 0.40$\\
$ \Xi_{c}^{+} \to \Sigma^{+} \bar{K}^{*0} $&$ 2.88\pm 1.06 $~\cite{pdg,Exp:absXicp}&$1.40\pm 0.69$&$ \Xi_{c}^{0} \to \Lambda^{0} \phi $&$ 0.49\pm 0.15 $~\cite{pdg}&$0.44\pm 0.08$\\
$ \Xi_{c}^{+} \to \Xi^{0} \rho^{+} $&$ 8.2\pm 3.6   $~\cite{pdg,Exp:absXicp}&$14.48\pm 2.44$&&\\
\hline
		\end{tabular}
	\end{center}
\end{table}

In our numerical calculations, we adopt the minimal $\chi^2$ fitting. We find that the minimal value of $\chi^2$/(degree of freedom) is
given by $18/4=4.5$, which is reasonable  as $SU(3)_F$ is not an exact symmetry.
The results of the effective coupling parameters are found to be
\begin{eqnarray}
(a_1,a_2,a_3,\tilde{a})&=&  ( -2.40\pm 0.24,0.82\pm 0.44,-2.05\pm 0.38,-1.59\pm 0.10      )   G_F \text{GeV}^2      \,,   \\
(b_1,b_2,b_3,\tilde{b})&=& ( 6.91\pm 0.28 , -0.82\pm 0.99, 2.82\pm 0.52, 0.75\pm 0.42 ) G_F \text{GeV}^2\,,
\end{eqnarray}
with the correlation in the sequences $(a_1,a_2,a_3,\tilde{a},b_1,b_2,b_3,\tilde{b})$, given by
\begin{eqnarray}
R=\left(
\begin{array}{ccccccccc}
1 &-0.087 &0.085 &-0.043 &0.423 &0.161 &-0.091 &0.083 \\
-0.087 &1 &0.599 &0.325 &0.043 &0.540 &0.105 &0.363 \\
0.085 &0.599 &1 &-0.094 &0.126 &0.257 &0.346 &-0.096 \\
-0.043 &0.325 &-0.094 &1 &-0.011 &0.473 &-0.308 &0.640 \\
0.423 &0.043 &0.126 &-0.011 &1 &0.314 &0.135 &-0.150 \\
0.161 &0.540 &0.257 &0.473 &0.314 &1 &-0.112 &0.472 \\
-0.091 &0.105 &0.346 &-0.308 &0.135 &-0.112 &1 &-0.355 \\
0.083 &0.363 &-0.096 &0.640 &-0.150 &0.472 &-0.355 &1 \\
\end{array}
\right)\,,
\end{eqnarray}
where $\tilde{a}=a_0+(a_1+a_2-a_3)/3$ and  $\tilde{b}=b_0+(b_1+b_2-b_3)/3$.
Accordingly, the branching ratios, up-down asymmetries and longitudinal polarizations
for the Cabbibo allowed, singly suppressed and doubly suppressed decays of ${\bf B}_c \to {\bf B}_n V$
 are shown in Tables~\ref{BRCabA}, \ref{BRCabS} and \ref{BRCabDS}, respectively.
The reconstructed branching ratios are also listed in Table~\ref{Table1}. 
Most of the results are consistent with the experimental data except  ${\cal B}(\Lambda_c^+ \to \Sigma^+ K^{*0})$.

\begin{table}[h]
	\begin{center}
		\centering
		\caption{ Cabbibo allowed decays of ${\bf B}_c \to {\bf B}_n V$.}\label{BRCabA}
		\begin{tabular}[t]{lrrrr}
			\hline
			channel & \multicolumn{1}{c}{$10^2{\cal B}_{SU(3)}$} & \multicolumn{1}{c}{$\alpha$} &\multicolumn{1}{c}{$P_L$} &\multicolumn{1}{c}{$\alpha_V$} \\
			\hline
 $ \Lambda_{c}^{+}  \to  \Lambda^{0} \rho^{+} $&$ 4.81 \pm 0.58^a $&$ -0.27 \pm 0.04 $&$ -0.93 \pm 0.05 $&$ 2.01 \pm 0.39 $\\
$ \Lambda_{c}^{+}  \to  p \bar{K}^{*0} $&$ 2.03 \pm 0.25 ^a$&$ -0.18 \pm 0.05 $&$ -0.62 \pm 0.16 $&$ 4.96 \pm 0.76 $\\
$ \Lambda_{c}^{+}  \to  \Sigma^{0} \rho^{+} $&$ 1.43 \pm 0.42 $&$ -0.34 \pm 0.18 $&$ -0.66 \pm 0.34 $&$ 9.82 \pm 7.19 $\\
$ \Lambda_{c}^{+}  \to  \Sigma^{+} \rho^{0} $&$ 1.43 \pm 0.42 ^a$&$ -0.34 \pm 0.18 $&$ -0.66 \pm 0.34 $&$ 9.82 \pm 7.19 $\\
$ \Lambda_{c}^{+}  \to  \Sigma^{+} \omega $&$ 1.81 \pm 0.19 ^a$&$ -0.34 \pm 0.11 $&$ -0.67 \pm 0.22 $&$ 1.60 \pm 0.62 $\\
$ \Lambda_{c}^{+}  \to  \Sigma^{+} \phi $&$ 0.39 \pm 0.06 ^a$&$ 0.02 \pm 0.03 $&$ 0.08 \pm 0.10 $&$  0.16 \pm 0.01 $\\
$ \Lambda_{c}^{+}  \to  \Xi^{0} K^{*+} $&$ 0.10 \pm 0.10 $&$ -0.15 \pm 0.20 $&$ -0.40 \pm 0.55 $&$  0.35 \pm 0.52 $\\
\hline
$ \Xi_{c}^{+}  \to  \Sigma^{+} \bar{K}^{*0} $&$ 1.40 \pm 0.69 ^a$&$ 0.32 \pm 0.30 $&$ 0.37 \pm 0.35 $&$40.30 ^{+ 68.54}_{-41.30}$\\
$ \Xi_{c}^{+}  \to  \Xi^{0} \rho^{+} $&$ 14.48 \pm 2.44 ^a$&$ 0.00 \pm 0.07 $&$ -0.62 \pm 0.13 $&$  1.07 \pm 0.09  $\\
\hline
$ \Xi_{c}^{0}  \to  \Lambda^{0} \bar{K}^{*0} $&$ 1.37 \pm 0.26 $&$ -0.28 \pm 0.10 $&$ -0.67 \pm 0.24 $&$6.94 \pm 2.28 $\\
$ \Xi_{c}^{0}  \to  \Sigma^{0} \bar{K}^{*0} $&$ 0.42 \pm 0.23 $&$ -0.33 \pm 0.50 $&$ -0.42 \pm 0.62 $&$ 38.99 ^{+ 82.32}_{-39.99} $\\
$ \Xi_{c}^{0}  \to  \Sigma^{+} K^{*-} $&$ 0.24 \pm 0.17 $&$ -0.37 \pm 0.31 $&$ -0.76 ^{+0.64}_{-0.24} $&$1.94 \pm 2.63 $\\
$ \Xi_{c}^{0}  \to  \Xi^{0} \rho^{0} $&$ 0.88 \pm 0.22 $&$ -0.15 \pm 0.18 $&$ -0.26 \pm 0.32 $&$ 20.55 \pm 5.91  $\\
$ \Xi_{c}^{0}  \to  \Xi^{0} \omega $&$ 2.78 \pm 0.45 $&$ -0.40 \pm 0.07 $&$ -0.71 \pm 0.12 $&$ 2.03 \pm 0.47 $\\
$ \Xi_{c}^{0}  \to  \Xi^{0} \phi $&$ 0.14 \pm 0.13 $&$ 0.22 \pm 0.10 $&$ 0.61 \pm 0.27 $&$ 0.71 \pm 0.51 $\\
$ \Xi_{c}^{0}  \to  \Xi^{-} \rho^{+} $&$ 8.98 \pm 0.55 $&$ -0.32 \pm 0.01 $&$ -0.94 \pm 0.01 $&$ 2.45 \pm 0.21$\\
			\hline
			\multicolumn{4}{l}{$^a$ reconstructed values}
		\end{tabular}
	\end{center}
\end{table}
\begin{table}
	\begin{center}
		\centering
		\caption{  Singly Cabbibo suppressed decays of ${\bf B}_c \to {\bf B}_n V$.}\label{BRCabS}
		\begin{tabular}[t]{lrrrr}
			\hline
			channel& \multicolumn{1}{c}{$10^3{\cal B}_{SU(3)}$} & \multicolumn{1}{c}{$\alpha$} &\multicolumn{1}{c}{$P_L$}&\multicolumn{1}{c}{$\alpha_V$}  \\
			\hline
$ \Lambda_{c}^{+}  \to  \Lambda^{0} K^{*+} $&$ 3.35 \pm 0.37 $&$ -0.13 \pm 0.05 $&$ -0.81 \pm 0.09 $&$ 1.03 \pm 0.22 $\\
$ \Lambda_{c}^{+}  \to  p \rho^{0} $&$ 0.02 ^{+ 0.07}_{-0.02} $&$ -0.27 ^{+1.27}_{-0.73} $&$ -0.28 ^{+1.28}_{-0.72} $&$ -$\\
$ \Lambda_{c}^{+}  \to  p \omega $&$ 0.63 \pm 0.34^a $&$ 0.36 \pm 0.17 $&$ 0.88 ^{+0.12}_{-0.21} $&$ 2.95 \pm 1.01 $\\
$ \Lambda_{c}^{+}  \to  p \phi $&$ 0.87 \pm 0.14 ^a$&$ -0.06 \pm 0.04 $&$ -0.83 \pm 0.08 $&$ 0.88 \pm 0.14 $\\
$ \Lambda_{c}^{+}  \to  n \rho^{+} $&$ 1.76 \pm 0.72 $&$ -0.09 \pm 0.22 $&$ -0.84 ^{+ 0.35}_{-0.16} $&$ 1.48 \pm 0.47 $\\
$ \Lambda_{c}^{+}  \to  \Sigma^{0} K^{*+} $&$ 0.18 \pm 0.04 $&$ -0.14 \pm 0.17 $&$ -0.35 \pm 0.41 $&$ 11.71 \pm 5.00 $\\
$ \Lambda_{c}^{+}  \to  \Sigma^{+} K^{*0} $&$ 0.38 \pm 0.09^a $&$ -0.14 \pm 0.17 $&$ -0.34 \pm 0.41 $&$ 11.86 \pm 4.99 $\\
\hline
$ \Xi_{c}^{+}  \to  \Lambda^{0} \rho^{+} $&$ 1.52 \pm 0.57 $&$ 0.49 \pm 0.22 $&$ 0.28 \pm 0.46 $&$ 2.05 \pm 0.82 $\\
$ \Xi_{c}^{+}  \to  p \bar{K}^{*0} $&$ 4.71 \pm 1.22 ^a$&$ -0.12 \pm 0.15 $&$ -0.23 \pm 0.29 $&$ 13.01 \pm 1.39 $\\
$ \Xi_{c}^{+}  \to  \Sigma^{0} \rho^{+} $&$ 11.45 \pm 1.52 $&$ -0.39 \pm 0.02 $&$ -0.96 \pm 0.00 $&$ 3.32 \pm 0.67 $\\
$ \Xi_{c}^{+}  \to  \Sigma^{+} \rho^{0} $&$ 2.85 \pm 0.81 $&$ -0.42 \pm 0.04 $&$ -0.91 ^{+ 0.12}_{-0.09} $&$ 4.99 \pm 2.14 $\\
$ \Xi_{c}^{+}  \to  \Sigma^{+} \omega $&$ 4.11 \pm 0.77 $&$ -0.13 \pm 0.17 $&$ -0.48 \pm 0.28 $&$ 1.68 \pm 0.23 $\\
$ \Xi_{c}^{+}  \to  \Sigma^{+} \phi $&$ 1.82 \pm 0.40^a $&$ -0.56 \pm 0.02 $&$ -0.75 \pm 0.06 $&$ 3.19 \pm 1.56 $\\
$ \Xi_{c}^{+}  \to  \Xi^{0} K^{*+} $&$ 4.28 \pm 1.64 $&$ 0.28 \pm 0.10 $&$ -0.45 \pm 0.27 $&$ 0.40 \pm 0.07 $\\
\hline
$ \Xi_{c}^{0}  \to  \Lambda^{0} \rho^{0} $&$ 0.13 \pm 0.11 $&$ 0.51 \pm 0.17 $&$ 0.72 \pm 0.21 $&$ 13.22 \pm 7.92 $\\
$ \Xi_{c}^{0}  \to  \Lambda^{0} \omega $&$ 1.51 \pm 0.20 $&$ -0.16 \pm 0.19 $&$ -0.19 \pm 0.31 $&$ 2.12 \pm 0.19 $\\
$ \Xi_{c}^{0}  \to  \Lambda^{0} \phi $&$ 0.44 \pm 0.08 ^a$&$ -0.10 \pm 0.13 $&$ -0.63 \pm 0.32 $&$ 0.90 \pm 0.36 $\\
$ \Xi_{c}^{0}  \to  p K^{*-} $&$ 0.19 \pm 0.14 $&$ -0.47 \pm 0.26 $&$ -0.88 ^{+ 0.49}_{-0.12} $&$ 3.36 \pm 3.92 $\\
$ \Xi_{c}^{0}  \to  n \bar{K}^{*0} $&$ 2.52 \pm 0.79 $&$ -0.31 \pm 0.19 $&$ -0.58 \pm 0.36 $&$ 10.29 \pm 3.73 $\\
$ \Xi_{c}^{0}  \to  \Sigma^{0} \rho^{0} $&$ 0.11 \pm 0.10 $&$ -0.08 \pm 0.25 $&$ -0.28 \pm 0.69 $&$ 6.42 \pm 4.14 $\\
$ \Xi_{c}^{0}  \to  \Sigma^{0} \omega $&$ 0.70 \pm 0.13 $&$ -0.13 \pm 0.17 $&$ -0.48 \pm 0.28 $&$ 1.70 \pm 0.24 $\\
$ \Xi_{c}^{0}  \to  \Sigma^{0} \phi $&$ 0.30 \pm 0.07 $&$ -0.57 \pm 0.02 $&$ -0.75 \pm 0.06 $&$ 3.21 \pm 1.57 $\\
$ \Xi_{c}^{0}  \to  \Sigma^{+} \rho^{-} $&$ 0.19 \pm 0.13 $&$ -0.50 \pm 0.32 $&$ -0.83 ^{+ 0.53}_{-0.17} $&$ 3.24 \pm 4.27 $\\
$ \Xi_{c}^{0}  \to  \Sigma^{-} \rho^{+} $&$ 5.56 \pm 0.34 $&$ -0.37 \pm 0.01 $&$ -0.97 \pm 0.01 $&$ 3.32 \pm 0.27 $\\
$ \Xi_{c}^{0}  \to  \Xi^{0} K^{*0} $&$ 0.79 \pm 0.23 $&$ -0.33 \pm 0.15 $&$ -0.71 ^{+ 0.32}_{-0.29} $&$ 7.39 \pm 6.19 $\\
$ \Xi_{c}^{0}  \to  \Xi^{-} K^{*+} $&$ 3.36 \pm 0.23 $&$ -0.12 \pm 0.01 $&$ -0.87 \pm 0.03 $&$ 1.07 \pm 0.09 $\\
			\hline
			\multicolumn{4}{l}{$^a$ reconstructed values}
		\end{tabular}
	\end{center}
\end{table}
\begin{table}[h]
	\begin{center}
		\centering
		\caption{Doubly Cabbibo  suppressed decays of ${\bf B}_c \to {\bf B}_n V$.}\label{BRCabDS}
		\begin{tabular}[t]{lrrrr}
			\hline
			channel & \multicolumn{1}{c}{$10^4{\cal B}_{SU(3)}$} & \multicolumn{1}{c}{$\alpha$} &\multicolumn{1}{c}{$P_L$} &\multicolumn{1}{c}{$\alpha_V$} \\
			\hline
$ \Lambda_{c}^{+}  \to  p K^{*0} $&$ 0.04 ^{+ 0.05}_{-0.04} $&$ 0.83 ^{+0.17}_{-0.21} $&$ 0.38 \pm 0.55 $&$ 2.13^{+ 3.35}_{-3.13} $\\
$ \Lambda_{c}^{+}  \to  n K^{*+} $&$ 1.48 \pm 0.25 $&$ -0.01 \pm 0.06 $&$ -0.78 \pm 0.12 $&$ 0.99 \pm 0.11 $\\
\hline
$ \Xi_{c}^{+}  \to  \Lambda^{0} K^{*+} $&$ 0.34 ^{+ 0.37}_{-0.34} $&$ 0.59 \pm 0.27 $&$ 0.62 \pm 0.31 $&$ 10.37^{+ 14.78}_{-11.37} $\\
$ \Xi_{c}^{+}  \to  p \rho^{0} $&$ 0.22 \pm 0.17 $&$ -0.58 \pm 0.25 $&$ -0.92 ^{+ 0.39}_{-0.08} $&$ 5.07 \pm 5.72 $\\
$ \Xi_{c}^{+}  \to  p \omega $&$ 1.66 \pm 0.70 $&$ 0.42 \pm 0.17 $&$ 0.69 \pm 0.28 $&$ 3.55 \pm 0.81 $\\
$ \Xi_{c}^{+}  \to  p \phi $&$ 2.29 \pm 0.39 $&$ -0.40 \pm 0.01 $&$ -0.93 \pm 0.03 $&$ 3.76 \pm 1.04 $\\
$ \Xi_{c}^{+}  \to  n \rho^{+} $&$ 0.43 \pm 0.33 $&$ -0.58 \pm 0.25 $&$ -0.92 ^{+ 0.39}_{-0.08}  $&$ 5.06 \pm 5.72 $\\
$ \Xi_{c}^{+}  \to  \Sigma^{0} K^{*+} $&$ 3.08 \pm 0.20 $&$ -0.16 \pm 0.01 $&$ -0.96 \pm 0.02 $&$ 1.56 \pm 0.12 $\\
$ \Xi_{c}^{+}  \to  \Sigma^{+} K^{*0} $&$ 0.40 \pm 0.08 $&$ -0.11 \pm 0.03 $&$ -0.89 ^{+0.16}_{-0.11} $&$ 1.97 \pm 0.63 $\\
\hline
$ \Xi_{c}^{0}  \to  \Lambda^{0} K^{*0} $&$ 0.28 \pm 0.13 $&$ -0.35 \pm 0.31 $&$ -0.58 ^{+ 0.51}_{-0.42} $&$ 14.64 \pm 13.20 $\\
$ \Xi_{c}^{0}  \to  p \rho^{-} $&$ 0.15 \pm 0.11 $&$ -0.58 \pm 0.25 $&$ -0.92 ^{+0.39}_{-0.08} $&$ 5.09 \pm 5.74 $\\
$ \Xi_{c}^{0}  \to  n \rho^{0} $&$ 0.07 \pm 0.06 $&$ -0.58 \pm 0.25 $&$ -0.92 ^{+0.39}_{-0.08} $&$ 5.08 \pm 5.74 $\\
$ \Xi_{c}^{0}  \to  n \omega $&$ 0.56 \pm 0.24 $&$ 0.43 \pm 0.17 $&$ 0.69 \pm 0.28 $&$ 3.56 \pm 0.82 $\\
$ \Xi_{c}^{0}  \to  n \phi $&$ 0.77 \pm 0.13 $&$ -0.40 \pm 0.01 $&$ -0.93 \pm 0.03 $&$ 3.77 \pm 1.04 $\\
$ \Xi_{c}^{0}  \to  \Sigma^{0} K^{*0} $&$ 0.07 \pm 0.01 $&$ -0.12 \pm 0.03 $&$  -0.89 ^{+0.16}_{-0.11}  $&$ 2.03 \pm 0.64 $\\
$ \Xi_{c}^{0}  \to  \Sigma^{-} K^{*+} $&$ 2.08 \pm 0.14 $&$ -0.17 \pm 0.01 $&$ -0.96 \pm 0.02 $&$ 1.58 \pm 0.12 $\\
			\hline
		\end{tabular}
	\end{center}
\end{table}

Note that our result of ${\cal B}(\Lambda_c^+ \to \Lambda^0 \rho^+)=(4.81\pm0.58)\times 10^{-2}$
agrees with  the  experimental upper limit of $6\times 10^{-2}$ (90\% C.L.)~\cite{pdg,LcLrhoold},
which is
obtained from
 ${\cal B}(\Lambda_c^+ \to \Lambda^0 \pi^+ \pi^0)/{\cal B}(\Lambda_c^+ \to p K^-\pi^+)<0.95$ by CLEO~\cite{LcLrhoold},
   where the resonant contribution of $ {\cal B}(\Lambda_c^+ \to\Lambda^0\rho^+, \rho^+\to \pi^+ \pi^0)$ is included in ${\cal B}(\Lambda_c^+ \to \Lambda^0 \pi^+ \pi^0)$ along with ${\cal B}(\Lambda_c^+ \to p K^-\pi^+)= (6.28\pm 0.32)\%$.
 However, it is inconsistent with the latest experimental measurement of  ${\cal B}(\Lambda_c^+ \to \Lambda^0 \pi^+ \pi^0)/{\cal B}(\Lambda_c^+ \to p K^-\pi^+)=1.20\pm 0.11$~\cite{Ablikim:2015flg}, making the experimental upper limit for ${\cal B}(\Lambda_c^+ \to \Lambda^0 \rho^+)$ questionable.

The other possible resonant dominated  contribution in $\Lambda_c^+\to \Lambda^0 \pi^+\pi^0$ is given by
\begin{eqnarray}
{\cal B}(\Lambda_c^+\to \Sigma(1385)^{*+(0)} \pi^{0(+)}, ~\Sigma^{*+(0)}\to \Lambda^0 \pi^{+(0)}) = (1.9\pm 0.4)\times 10^{-3}\,,
\end{eqnarray}
where we have taken ${\cal B}(\Lambda_c^+\to \Sigma(1385) \pi)=(2.2\pm 0.4)\times 10^{-3}$ from our previous work~\cite{SU(3)Dec} and ${\cal B}(\Sigma(1385) \to \Lambda\pi) = 0.87\pm 0.01$ in PDG~\cite{pdg}.
By subtracting these resonant contributions in $\Lambda_c^+\to \Lambda\pi^+\pi^0$, we find that
\begin{eqnarray}
{\cal B}(\Lambda_c^+\to \Lambda^0\pi^0\pi^+)_{non}/{\cal B}(\Lambda_c^+\to \Lambda\pi^0\pi^+) < 44\%
\end{eqnarray}
with $90\%$ C.L.
and ${\cal B}(\Lambda_c^+\to \Lambda^0\pi^0\pi^+)_{non}= (1.9\pm 0.7)\%$ by neglecting other resonant channels, 
where the subscript of $``non''$ represents the non-resonant contribution only.
This result shows   that $\Lambda_c^+ \to \Lambda^0 \pi^+\pi^0$ is dominated by the resonances,  which is one of the important predictions in  Ref.~\cite{Latest_three}.

The decays of $\Lambda_c^+\to \Sigma^+K^{*0}$ and $\Xi_c^+ \to p \bar{K}^{*0}$ share the same coupling strengths in
terms of the $U-$spin symmetry~\cite{Jia:2019zxi} as they are related through interchanging $d$ and $s$ quarks. 
Naively, one  expects that  they should have the same decay widths. However, our results indicate that
\begin{equation}\label{Ratio}
\Gamma(\Lambda_c^+ \to\Sigma^+ K^{*0})/\Gamma(\Xi_c^+ \to p \bar{K}^{*0})_{SU(3)_F}=0.18\pm 0.01\,.
\end{equation}
This hierarchy can be understood by the released energies, given by
\begin{eqnarray}
m_{\Lambda_c^+}-m_{\Sigma^+}-m_{K^*0} &=& 0.20 \text{GeV}\nonumber\\
m_{\Xi_c^+} - m_p - m_{\bar{K}^{*0}} &= & 0. 64 \text{GeV}\,.
\end{eqnarray}
With a smaller kinematic phase space, the decay of $\Lambda_c^+ \to \Sigma^+ K^{*0}$ is suppressed compared to $\Xi_c^+ \to p\bar{K}^{*0}$. 
It can be interpreted as the $SU(3)_F$ breaking effect, caused by the mass differences.
Meanwhile, the experimental data lead to
\begin{equation}
\Gamma(\Lambda_c^+ \to\Sigma^+ K^{*0})/\Gamma(\Xi_c^+ \to p \bar{K}^{*0})_{ex}=1.9\pm 0.8\,,
\end{equation}
which is much larger than the value in Eq.~(\ref{Ratio}).
Despite this inconsistence,   we are still confident that our result in Eq.~(\ref{Ratio}) due to the phase space suppression is correct.
 We view this result  as one of our predictions and suggest the future experiments to revisit the channels.

It is interesting to note that
the Cabbibo allowed decays of $\Lambda_c^+ \to \Lambda^0 \rho^+$ and $ \Xi_c^0 \to \Xi^- \rho^+$ 
have large branching ratios and decay parameters with small uncertainties
as shown in Table~\ref{BRCabA},
so that they can be viewed as the golden modes for the experimental searches.
Similarly, the singly Cabbibo  suppressed decays of  $\Lambda_c^+ \to \Lambda^0 K^{*+}$, $\Xi_c^+ \to \Sigma^+ \phi$ and $\Xi_c^0\to \Xi^- K^{*+}$
 are also recommended to  future experiments  for the same reasons. In addition, we point out that the decay parameters
  in  $\Xi_c^{+(0)}\to \Sigma^{+(0)} \phi$ are almost the same in terms of the isospin symmetry. However, the decay branching ratio 
  for the neutral  $\Xi_c^0$ mode is suppressed due to the shorter lifetime compared to the $\Xi_c^+$ one and the factor 2 from the CG coefficient.

In Table~\ref{Compare}, we compare our results of  the Cabbibo allowed decays   with those in the literature, where
 K$\ddot{\text{o}}$rner and Kr$\ddot{\text{a}}$mer (KK)~\cite{Korner:1992wi}, $\dot{\text{Z}}$enczykowski (Zen)~\cite{Zenczykowski:1993jm}
 and Hsiao, Yu and Zhao (HYZ)~\cite{Hsiao:2019yur} are the studies based on
 the covariant quark model,   pole model and  $SU(3)_F$, respectively.
In Ref.~\cite{Korner:1992wi}, only the decay widths are provided instead of the branching ratios.
To obtain the branching ratios, we have used the lifetimes in Eq.~(\ref{times}). 
 As seen from Table~\ref{Compare}, our results are consistent with those in 
 Ref.~\cite{Korner:1992wi}.
 However,  
 the branching ratios of
  $\rho^+$ modes of  $\Lambda_c^+\to \Lambda^0\rho^+$, $\Lambda_c^+\to \Sigma^0\rho^+$, $\Xi_c^+\to \Xi^0\rho^+$ and
   $\Xi_c^0\to \Xi^-\rho^+$ in Ref.~\cite{Korner:1992wi} are too large
   compared to our predictions as well as  the existing data.
  Furthermore, most of our results are compatible with those in Ref.~\cite{Zenczykowski:1993jm},  
 whereas differ largely in $\Lambda_c^+ \to \Lambda^0 \rho^+$, $\Xi_c^0\to( \Xi^- \rho^+,\Xi^0\omega)$ and  $\Xi_c^+$ decays. 
 Note that in Ref.~\cite{Hsiao:2019yur}, the contributions from $H(\overline{15})$ and the correlations between the parameters
  are not included in their calculations,
resulting in larger errors than ours.
 Except  the decays with the existing experimental data,
 which are also the inputs for the fitting, the predictions in Ref.~\cite{Hsiao:2019yur} 
 are quite different from ours even though both of us take the $SU(3)_F$ approach.
 In particular, due to the different treatments of the wave amplitude,
 the predicted decay branching ratio of $\Lambda_c^+ \to \Xi^0K^{*+}$ in Ref.~\cite{Hsiao:2019yur} is
  about 8 times larger than ours and the one in the literature~\cite{Korner:1992wi,Zenczykowski:1993jm}. 
\begin{table}
\caption{ Decay branching ratios~(\%) of the Cabbibo favored channels in our  $SU(3)_F$ approach and those in 
		K$\ddot{\text{o}}$rner and Kr$\ddot{\text{a}}$merl (KK)~\cite{Korner:1992wi}, $\dot{\text{Z}}$enczykowski (Zen)~\cite{Zenczykowski:1993jm}
 and Hsiao, Yu and Zhao (HYZ)~\cite{Hsiao:2019yur} along with the data in Ref.~\cite{pdg}.}
\label{Compare}
\begin{center}
		\begin{tabular}[t]{lcccccc}
			\hline
			channel & Our results&KK~\cite{Korner:1992wi}&Zen~\cite{Zenczykowski:1993jm}&HYZ~\cite{Hsiao:2019yur}&Data~\cite{pdg}\\ 
				\hline
$ \Lambda_{c}^{+}  \to  \Lambda^{0} \rho^{+} $&$ 4.81 \pm 0.58 $&$19.4$&1.80&$0.74\pm 0.34$&$<6$\\
$ \Lambda_{c}^{+}  \to  p \bar{K}^{*0} $&$ 2.03 \pm 0.25 $&3.13&5.03&$1.9\pm 0.3$&$1.96\pm 0.27$\\
$ \Lambda_{c}^{+}  \to  \Sigma^{0} \rho^{+} $&$ 1.43 \pm 0.42 $&3.19&1.56&$0.61\pm 0.46 $& \\
$ \Lambda_{c}^{+}  \to  \Sigma^{+} \rho^{0} $&$ 1.43 \pm 0.42 $&3.17&1.56&$0.61\pm 0.46 $&\\
$ \Lambda_{c}^{+}  \to  \Sigma^{+} \omega $&$ 1.81 \pm 0.19 $&4.09&1.10&$1.6\pm 0.7$&$1.70\pm 0.21$\\
$ \Lambda_{c}^{+}  \to  \Sigma^{+} \phi $&$ 0.39 \pm 0.06 $&0.26&0.11&$0.39\pm 0.06$&$0.39\pm 0.06$\\
$ \Lambda_{c}^{+}  \to  \Xi^{0} K^{*+} $&$ 0.10 \pm 0.10 $&0.12&0.11&$0.87\pm 0.27$\\
\hline
$ \Xi_{c}^{+}  \to  \Sigma^{+} \bar{K}^{*0} $&$ 1.40 \pm 0.69 $&2.42&7.38&$10.1\pm 2.9$&$2.88\pm 1.06$\\
$ \Xi_{c}^{+}  \to  \Xi^{0} \rho^{+} $&$ 14.48 \pm 2.44 $&99.0&5.48&$9.9\pm 2.9$&$8.2\pm 3.6$\\
\hline
$ \Xi_{c}^{0}  \to  \Lambda^{0} \bar{K}^{*0} $&$ 1.37 \pm 0.26 $&1.55&1.15&$0.46\pm 0.21$\\
$ \Xi_{c}^{0}  \to  \Sigma^{0} \bar{K}^{*0} $&$ 0.42 \pm 0.23 $&0.85&0.77&$0.27\pm 0.22$\\
$ \Xi_{c}^{0}  \to  \Sigma^{+} K^{*-} $&$ 0.24 \pm 0.17 $&0.54&0.37&$0.93\pm 0.29$\\
$ \Xi_{c}^{0}  \to  \Xi^{0} \rho^{0} $&$ 0.88 \pm 0.22 $&2.36&1.22&$1.4\pm 0.4 $\\
$ \Xi_{c}^{0}  \to  \Xi^{0} \omega $&$ 2.78 \pm 0.45 $&3.21&0.15&$0.10^{+0.86}_{-0.10}$\\
$ \Xi_{c}^{0}  \to  \Xi^{0} \phi $&$ 0.14 \pm 0.13 $&0.25&0.10&$0.015^{+0.071}_{-0.015}$\\
$ \Xi_{c}^{0}  \to  \Xi^{-} \rho^{+} $&$ 8.98 \pm 0.55 $&16.9&1.50&$0.86\pm 0.12$\\
			\hline
		\end{tabular}
	\end{center}
\end{table}

\section{Conclusions}
We have explored the charmed baryon decays of ${\bf B}_c\to {\bf B}_n V$ based on the $SU(3)_F$ flavor symmetry. 
In these processes, we have calculated  the color-symmetric parts of the effective Hamiltonian by the factorization approach assisted with 
the MIT bag model, while the anti-symmetric ones are extracted from the experimental data.
We have systematically obtained all decay branching ratios and   parameters in ${\bf B}_c\to {\bf B}_n V$. 
 We have found that our results are consistent with the experimental data  except  $\Lambda_c^+ \to \Sigma^0 K^{*+}$,
 for which our fitted value of ${\cal B}(\Lambda_c^+ \to \Sigma^0 K^{*+})=(0.38\pm 0.09)\times 10^{-3}$
 is much smaller than the data of $(3.5\pm 1.0)\times 10^{-3}$.
 As our result contains a very small error, whereas the experimental one is large, we are eager to see the precision measurement of this mode in the future
 experiments. 
 We have demonstrated  that the branching ratios of $\Lambda_c^+\to \Lambda^0\pi^+\pi^0$ and $\Xi_c^+\to \Xi^0\pi^+\pi^0$ are  dominated 
by the resonances with the decay chains of $\Lambda_c^+ \to \Lambda^0 \rho^+,~\rho^+\to \pi^+\pi^0$ and $\Xi_c^+ \to \Xi^0\rho^+,~\rho^+ \to \pi^+\pi^0$, respectively.
We have shown that most of our results with  $SU(3)_F$ are consistent with those 
 calculated from the dynamical models in Refs.~\cite{Korner:1992wi} and  \cite{Zenczykowski:1993jm}.
 However, the predictions for 
  the $\rho^+$ modes of  $\Lambda_c^+\to \Lambda^0\rho^+$, $\Lambda_c^+\to \Sigma^0\rho^+$, $\Xi_c^+\to \Xi^0\rho^+$ and
   $\Xi_c^0\to \Xi^-\rho^+$ in Ref.~\cite{Korner:1992wi} are too large, 
   whereas those of $\Lambda_c^+ \to \Lambda^0 \rho^+$ and  $\Xi_c^0\to (\Xi^- \rho^+,\Xi^0\omega)$ 
   in Ref.~\cite{Zenczykowski:1993jm} are found too small, compared to  our  values.
On the other hand,  our results are very different from those in Ref.~\cite{Hsiao:2019yur}, in which  the $SU(3)_F$
approach is also used but the contributions from color-symmetric parts of the effective Hamiltonian are ignored.

\appendix

\section{Dynamics}
To get a consistent results with the $SU(3)_F$ representation in Sec.~\MakeUppercase{\romannumeral 2}, we adopt the baryon wave functions as
\begin{equation}\label{B1}
{\bf B}_c = \frac{1}{\sqrt{6}}\left[  ({\bf B}_c)_{k}\epsilon^{ijk} q_iq_j c\otimes \chi_A +(23)+(13)            \right]
\end{equation}
 and
\begin{equation}\label{B2}
{\bf B}_n = \frac{1}{3} \left[ ({\bf B}_n)^i_l \epsilon^{ljk} q_jq_kq_i\otimes \chi_A   +(23)+(13)     \right]
\end{equation}
for the anti-triplet charmed and  octet baryons, respectively,
where $(23)$ stands for interchanging second and third quarks in the first term, while $(13)$  for  first and third ones. 
Here, the spin structure is defined as $\chi_A = (\uparrow \downarrow\uparrow-\downarrow\uparrow\uparrow)/\sqrt{2}$ .

The definitions in Eqs.~(\ref{B1}) and (\ref{B2}) have different signs for $\Xi^-$ and $\Lambda_c^+$ compared to those in
Refs.~\cite{Cheng:sup,Cheng:latest}, while they differ in sign for $\Sigma^+$, $\Xi^0$ and $\Lambda^0$ in  Ref.~\cite{Mit}. 

In this work,  the form factors are evaluated from the  MIT bag model~\cite{BAG,Mit}.
We follow the calculations in Ref.~\cite{Mit}. 
For completeness, the input parameters  are given by
\begin{equation}
m_{u,d}= 0.005~\text{GeV}\,,~~~~m_s= 0.28~\text{GeV}\,,~~~~m_c = 1.5~ \text{GeV}\,,~~~~R=5~\text{GeV}^{-1}\,,
\end{equation}
where $R$ is the radius of the quark bag.
After correcting a typo in the original derivation, Eq.~(19f) in Ref.~\cite{Mit} shall be read as
\begin{equation}
{\cal A}_T = (A-B) N^iN^fR^3W^i_-W^f_-J_{11}(-2R^2/15)\,,
\end{equation}
where ${\cal A}_T$ is one of the components in the axial vector, $A(B)$ is the quark overlapping factor for  the spin up (down), 
$N^i(N^f)$ is the normalized factor for the initial (final) baryon, $W^{i(f)}_-$ is associated with the normalized factor for quarks 
and $J_{11}$ is the overlapping between two Bessel functions.
The details can be found in Ref.~\cite{Mit}. 

Our results with $q^2 =0 $ are provided in Table~\ref{Table5}, where we have assumed the dipole momentum dependences, given by
\begin{eqnarray}
f_i = \frac{f(0)}{1-\frac{q^2}{M_V^2}}\,,~~~~
g_i=\frac{g(0)}{1-\frac{q^2}{M_A^2}}\,,
\end{eqnarray}
with $(M_V, M_A) = (2.01\,, 2.42)$~GeV for $c\to s$ and $(M_V, M_A) = (2.11\,, 2.51)$~GeV for $c\to u/d$. The sign differences in the 
form factors compared to Ref.~\cite{Mit} are due to the baryon wave functions.

\begin{table}[h]
	\begin{center}
		\centering
		\caption{Form factors for charmed baryons decaying to octet baryons with $q^2=0$.}\label{Table5}
		\begin{tabular}[t]{lrrrrrr}
			\hline
			channel & \multicolumn{1}{c}{$f_1$} & \multicolumn{1}{c}{$f_2$} &\multicolumn{1}{c}{$f_3$} &\multicolumn{1}{c}{$g_1$} &\multicolumn{1}{c}{$g_2$} &\multicolumn{1}{c}{$g_3$} \\
			\hline
			$ \Lambda_{c}^{+} \to \Lambda^{0} $&$ -0.455 $&$ -0.189 $&$ -0.001 $&$ -0.497 $&$ 0.055 $&$ 0.438 $\\
			$ \Lambda_{c}^{+} \to p $&$ 0.328 $&$ 0.181 $&$ 0.000 $&$ 0.407 $&$ -0.070 $&$ -0.501 $\\
			$ \Lambda_{c}^{+} \to n $&$ 0.330 $&$ 0.182 $&$ -0.000 $&$ 0.408 $&$ -0.070 $&$ -0.502 $\\
			\hline
			$ \Xi_{c}^{+} \to \Lambda^{0} $&$ -0.138 $&$ -0.093 $&$ 0.009 $&$ -0.168 $&$ 0.026 $&$ 0.271 $\\
			$ \Xi_{c}^{+} \to \Sigma^{0} $&$ 0.290 $&$ 0.201 $&$ -0.031 $&$ 0.332 $&$ -0.031 $&$ -0.550 $\\
			$ \Xi_{c}^{+} \to \Sigma^{+} $&$ -0.410 $&$ -0.285 $&$ 0.044 $&$ -0.469 $&$ 0.044 $&$ 0.778 $\\
			$ \Xi_{c}^{+} \to \Xi^{0} $&$ -0.587 $&$ -0.309 $&$ 0.029 $&$ -0.630 $&$ 0.053 $&$ 0.732 $\\
			\hline
			$ \Xi_{c}^{0} \to \Lambda^{0} $&$ 0.137 $&$ 0.093 $&$ -0.009 $&$ 0.167 $&$ -0.026 $&$ -0.271 $\\
			$ \Xi_{c}^{0} \to \Sigma^{0} $&$ 0.288 $&$ 0.201 $&$ -0.031 $&$ 0.330 $&$ -0.031 $&$ -0.549 $\\
			$ \Xi_{c}^{0} \to \Sigma^{-} $&$ 0.408 $&$ 0.284 $&$ -0.044 $&$ 0.467 $&$ -0.045 $&$ -0.777 $\\
			$ \Xi_{c}^{0} \to \Xi^{-} $&$ 0.590 $&$ 0.312 $&$ -0.030 $&$ 0.632 $&$ -0.052 $&$ -0.738 $\\
			\hline
			
		\end{tabular}
	\end{center}
\end{table}

\section{Amplitudes with $SU(3)_F$ representations}

In this Appendix, we provide the effective coupling strengths in Tables~\ref{AmpCA}, \ref{AmpCS} and \ref{AmpCDS}.
We distinguish $A_1(B_1)$ in two different parts according to the effective Hamiltonian. 
$A_1^{(\overline{15})}$ and $B_1^{(\overline{15})}$ are purely factorizable, which are calculated through the form factors.  
On the other hand, $A_1^{(6)}$ is parametrized by the $SU(3)_F$ symmetry, while   $B_1^{(6)}$ is  obtained by substituting $b_i$ for $a_i$. 
The contributions from $A_2$ and $B_2$ are suppressed as implied by Eq.~\eqref{A2supp}. We only consider the factorizable contributions in 
$A_2$ and $B_2$ for consistency with the calculation in $\Lambda_c^+\to p \phi$.
\begin{table}
	\begin{center}
		\centering
		\caption{Effective coupling strengths for Cabbibo allow decays with  units $10^{-1}~G_F$GeV$^2$.}\label{AmpCA}
		\begin{tabular}[t]{lcccccc}
			\hline
			channel & $A_{1}^{(\overline{15}) }$&$A_2^{(fac)}$& $B_{1}^{(\overline{15}) }$&$B_2^{(fac)}$&$A_1^{(6)}$\\
			\hline
$ \Lambda_{c}^{+}  \to  \Lambda^{0} \rho^{+} $&$ 0.281 $&$ 0.084 $&$ -0.430 $&$ 0.312 $&$ - \frac{\sqrt{6} a_{1}}{3} - \frac{\sqrt{6} a_{2}}{3} - \frac{\sqrt{6} a_{3}}{3} $\\
$ \Lambda_{c}^{+}  \to  p \bar{K}^{*0} $&$ -0.305 $&$ 0.052 $&$ 0.460 $&$ 0.154 $&$ - 2 a_{1} $\\
$ \Lambda_{c}^{+}  \to  \Sigma^{0} \rho^{+} $&$ 0 $&$ 0 $&$ 0 $&$ 0 $&$ - \sqrt{2} a_{1} + \sqrt{2} a_{2} + \sqrt{2} a_{3} $\\
$ \Lambda_{c}^{+}  \to  \Sigma^{+} \rho^{0} $&$ 0 $&$ 0 $&$ 0 $&$ 0 $&$ \sqrt{2} a_{1} - \sqrt{2} a_{2} - \sqrt{2} a_{3} $\\
$ \Lambda_{c}^{+}  \to  \Sigma^{+} \omega $&$ 0 $&$ 0 $&$ 0 $&$ 0 $&$ - 2 \sqrt{2} \tilde{a} - \frac{\sqrt{2} a_{1}}{3} - \frac{\sqrt{2} a_{2}}{3} + \frac{\sqrt{2} a_{3}}{3} $\\
$ \Lambda_{c}^{+}  \to  \Sigma^{+} \phi $&$ 0 $&$ 0 $&$ 0 $&$ 0 $&$ - 2 \tilde{a} + \frac{2 a_{1}}{3} + \frac{2 a_{2}}{3} - \frac{2 a_{3}}{3} $\\
$ \Lambda_{c}^{+}  \to  \Xi^{0} K^{*+} $&$ 0 $&$ 0 $&$ 0 $&$ 0 $&$ - 2 a_{2} $\\
\hline
$ \Xi_{c}^{+}  \to  \Sigma^{+} \bar{K}^{*0} $&$ 0.335 $&$ -0.030 $&$ -0.656 $&$ -0.224 $&$ 2 a_{3} $\\
$ \Xi_{c}^{+}  \to  \Xi^{0} \rho^{+} $&$ 0.350 $&$ 0.074 $&$ -0.619 $&$ 0.472 $&$ - 2 a_{3} $\\
\hline
$ \Xi_{c}^{0}  \to  \Lambda^{0} \bar{K}^{*0} $&$ -0.123 $&$ 0.018 $&$ 0.213 $&$ 0.073 $&$ - \frac{2 \sqrt{6} a_{1}}{3} + \frac{\sqrt{6} a_{2}}{3} + \frac{\sqrt{6} a_{3}}{3} $\\
$ \Xi_{c}^{0}  \to  \Sigma^{0} \bar{K}^{*0} $&$ -0.236 $&$ 0.021 $&$ 0.461 $&$ 0.158 $&$ - \sqrt{2} a_{2} - \sqrt{2} a_{3} $\\
$ \Xi_{c}^{0}  \to  \Sigma^{+} K^{*-} $&$ 0 $&$ 0 $&$ 0 $&$ 0 $&$ 2 a_{2} $\\
$ \Xi_{c}^{0}  \to  \Xi^{0} \rho^{0} $&$ 0 $&$ 0 $&$ 0 $&$ 0 $&$ - \sqrt{2} a_{1} + \sqrt{2} a_{3} $\\
$ \Xi_{c}^{0}  \to  \Xi^{0} \omega $&$ 0 $&$ 0 $&$ 0 $&$ 0 $&$ 2 \sqrt{2} \tilde{a} + \frac{\sqrt{2} a_{1}}{3} - \frac{2 \sqrt{2} a_{2}}{3} - \frac{\sqrt{2} a_{3}}{3} $\\
$ \Xi_{c}^{0}  \to  \Xi^{0} \phi $&$ 0 $&$ 0 $&$ 0 $&$ 0 $&$ 2 \tilde{a} - \frac{2 a_{1}}{3} + \frac{4 a_{2}}{3} + \frac{2 a_{3}}{3} $\\
$ \Xi_{c}^{0}  \to  \Xi^{-} \rho^{+} $&$ -0.351 $&$ -0.073 $&$ 0.624 $&$ -0.477 $&$ 2 a_{1} $\\
			\hline			
		\end{tabular}
	\end{center}
\end{table}
\begin{table}
	\begin{center}
		\centering
		\caption{Effective coupling strengths for the singly Cabbibo  suppressed decays with  units $10^{-2}~G_F$GeV$^2$.}\label{AmpCS}
		\begin{tabular}[t]{lcccccc}
			\hline
			channel & $A_{1}^{(\overline{15}) }$&$A_2^{(fac)}$& $B_{1}^{(\overline{15}) }$&$B_2^{(fac)}$&$s_c^{-1}A_1^{(6)}$\\
			\hline
$ \Lambda_{c}^{+}  \to  \Lambda^{0} K^{*+} $&$ 0.812 $&$ 0.242 $&$ -1.291 $&$ 0.937 $&$ - \frac{\sqrt{6} a_{1}}{3} + \frac{2 \sqrt{6} a_{2}}{3} - \frac{\sqrt{6} a_{3}}{3} $\\
$ \Lambda_{c}^{+}  \to  p \rho^{0} $&$ -0.397 $&$ 0.067 $&$ 0.574 $&$ 0.192 $&$ - \sqrt{2} a_{2} - \sqrt{2} a_{3} $\\
$ \Lambda_{c}^{+}  \to  p \omega $&$ 0.406 $&$ -0.069 $&$ -0.589 $&$ -0.197 $&$ - 2 \sqrt{2} \tilde{a} + \frac{2 \sqrt{2} a_{1}}{3} - \frac{\sqrt{2} a_{2}}{3} + \frac{\sqrt{2} a_{3}}{3} $\\
$ \Lambda_{c}^{+}  \to  p \phi $&$ -0.889 $&$ 0.150 $&$ 1.420 $&$ 0.476 $&$ - 2 \tilde{a} - \frac{4 a_{1}}{3} + \frac{2 a_{2}}{3} - \frac{2 a_{3}}{3} $\\
$ \Lambda_{c}^{+}  \to  n \rho^{+} $&$ 0.563 $&$ 0.248 $&$ -0.815 $&$ 0.716 $&$ - 2 a_{2} - 2 a_{3} $\\
$ \Lambda_{c}^{+}  \to  \Sigma^{0} K^{*+} $&$ 0 $&$ 0 $&$ 0 $&$ 0 $&$ - \sqrt{2} a_{1} + \sqrt{2} a_{3} $\\
$ \Lambda_{c}^{+}  \to  \Sigma^{+} K^{*0} $&$ 0 $&$ 0 $&$ 0 $&$ 0 $&$ - 2 a_{1} + 2 a_{3} $\\
\hline
$ \Xi_{c}^{+}  \to  \Lambda^{0} \rho^{+} $&$ -0.228 $&$ -0.085 $&$ 0.379 $&$ -0.339 $&$ - \frac{\sqrt{6} a_{1}}{3} - \frac{\sqrt{6} a_{2}}{3} + \frac{2 \sqrt{6} a_{3}}{3} $\\
$ \Xi_{c}^{+}  \to  p \bar{K}^{*0} $&$ 0 $&$ 0 $&$ 0 $&$ 0 $&$ - 2 a_{1} + 2 a_{3} $\\
$ \Xi_{c}^{+}  \to  \Sigma^{0} \rho^{+} $&$ 0.436 $&$ 0.102 $&$ -0.818 $&$ 0.733 $&$ - \sqrt{2} a_{1} + \sqrt{2} a_{2} $\\
$ \Xi_{c}^{+}  \to  \Sigma^{+} \rho^{0} $&$ 0.436 $&$ -0.039 $&$ -0.818 $&$ -0.280 $&$ \sqrt{2} a_{1} - \sqrt{2} a_{2} $\\
$ \Xi_{c}^{+}  \to  \Sigma^{+} \omega $&$ -0.446 $&$ 0.040 $&$ 0.840 $&$ 0.287 $&$ - 2 \sqrt{2} \tilde{a} - \frac{\sqrt{2} a_{1}}{3} - \frac{\sqrt{2} a_{2}}{3} - \frac{2 \sqrt{2} a_{3}}{3} $\\
$ \Xi_{c}^{+}  \to  \Sigma^{+} \phi $&$ 0.975 $&$ -0.087 $&$ -2.025 $&$ -0.692 $&$ - 2 \tilde{a} + \frac{2 a_{1}}{3} + \frac{2 a_{2}}{3} + \frac{4 a_{3}}{3} $\\
$ \Xi_{c}^{+}  \to  \Xi^{0} K^{*+} $&$ 1.011 $&$ 0.215 $&$ -1.858 $&$ 1.418 $&$ - 2 a_{2} - 2 a_{3} $\\
\hline
$ \Xi_{c}^{0}  \to  \Lambda^{0} \rho^{0} $&$ -0.161 $&$ 0.023 $&$ 0.266 $&$ 0.091 $&$ \frac{\sqrt{3} a_{1}}{3} + \frac{\sqrt{3} a_{2}}{3} - \frac{2 \sqrt{3} a_{3}}{3} $\\
$ \Xi_{c}^{0}  \to  \Lambda^{0} \omega $&$ 0.164 $&$ -0.024 $&$ -0.273 $&$ -0.093 $&$ - 2 \sqrt{3} \tilde{a} + \frac{\sqrt{3} a_{1}}{3} + \frac{\sqrt{3} a_{2}}{3} $\\
$ \Xi_{c}^{0}  \to  \Lambda^{0} \phi $&$ -0.359 $&$ 0.051 $&$ 0.658 $&$ 0.225 $&$ - \sqrt{6} \tilde{a} - \frac{\sqrt{6} a_{1}}{3} - \frac{\sqrt{6} a_{2}}{3} $\\
$ \Xi_{c}^{0}  \to  p K^{*-} $&$ 0 $&$ 0 $&$ 0 $&$ 0 $&$ 2 a_{2} $\\
$ \Xi_{c}^{0}  \to  n \bar{K}^{*0} $&$ 0 $&$ 0 $&$ 0 $&$ 0 $&$ - 2 a_{1} + 2 a_{2} + 2 a_{3} $\\
$ \Xi_{c}^{0}  \to  \Sigma^{0} \rho^{0} $&$ -0.307 $&$ 0.028 $&$ 0.575 $&$ 0.197 $&$ - a_{1} - a_{2} $\\
$ \Xi_{c}^{0}  \to  \Sigma^{0} \omega $&$ 0.314 $&$ -0.028 $&$ -0.591 $&$ -0.202 $&$ 2 \tilde{a} + \frac{a_{1}}{3} + \frac{a_{2}}{3} + \frac{2 a_{3}}{3} $\\
$ \Xi_{c}^{0}  \to  \Sigma^{0} \phi $&$ -0.687 $&$ 0.062 $&$ 1.423 $&$ 0.486 $&$ \sqrt{2} \tilde{a} - \frac{\sqrt{2} a_{1}}{3} - \frac{\sqrt{2} a_{2}}{3} - \frac{2 \sqrt{2} a_{3}}{3} $\\
$ \Xi_{c}^{0}  \to  \Sigma^{+} \rho^{-} $&$ 0 $&$ 0 $&$ 0 $&$ 0 $&$ - 2 a_{2} $\\
$ \Xi_{c}^{0}  \to  \Sigma^{-} \rho^{+} $&$ 0.614 $&$ 0.146 $&$ -1.151 $&$ 1.031 $&$ - 2 a_{1} $\\
$ \Xi_{c}^{0}  \to  \Xi^{0} K^{*0} $&$ 0 $&$ 0 $&$ 0 $&$ 0 $&$ 2 a_{1} - 2 a_{2} - 2 a_{3} $\\
$ \Xi_{c}^{0}  \to  \Xi^{-} K^{*+} $&$ -1.014 $&$ -0.210 $&$ 1.873 $&$ -1.431 $&$ 2 a_{1} $\\
			\hline			
		\end{tabular}
	\end{center}
\end{table}
\begin{table}
	\begin{center}
		\centering
		\caption{Effective coupling strengths for the doubly Cabbibo  suppressed decays with  units $10^{-3}~G_F$GeV$^2$.}\label{AmpCDS}
		\begin{tabular}[t]{lcccccc}
			\hline
			channel & $A_{1}^{(\overline{15}) }$&$A_2^{(fac)}$& $B_{1}^{(\overline{15}) }$&$B_2^{(fac)}$&$s_c^{-2}A_1^{(6)}$\\
			\hline
$ \Lambda_{c}^{+}  \to  p K^{*0} $&$ 1.623 $&$ -0.275 $&$ -2.449 $&$ -0.820 $&$ 2 a_{3} $\\
$ \Lambda_{c}^{+}  \to  n K^{*+} $&$ 1.638 $&$ 0.723 $&$ -2.481 $&$ 2.179 $&$ - 2 a_{3} $\\
\hline
$ \Xi_{c}^{+}  \to  \Lambda^{0} K^{*+} $&$ -0.664 $&$ -0.246 $&$ 1.154 $&$ -1.031 $&$ - \frac{\sqrt{6} a_{1}}{3} + \frac{2 \sqrt{6} a_{2}}{3} + \frac{2 \sqrt{6} a_{3}}{3} $\\
$ \Xi_{c}^{+}  \to  p \rho^{0} $&$ 0 $&$ 0 $&$ 0 $&$ 0 $&$ - \sqrt{2} a_{2} $\\
$ \Xi_{c}^{+}  \to  p \omega $&$ 0 $&$ 0 $&$ 0 $&$ 0 $&$ - 2 \sqrt{2} \tilde{a} + \frac{2 \sqrt{2} a_{1}}{3} - \frac{\sqrt{2} a_{2}}{3} - \frac{2 \sqrt{2} a_{3}}{3} $\\
$ \Xi_{c}^{+}  \to  p \phi $&$ 0 $&$ 0 $&$ 0 $&$ 0 $&$ - 2 \tilde{a} - \frac{4 a_{1}}{3} + \frac{2 a_{2}}{3} + \frac{4 a_{3}}{3} $\\
$ \Xi_{c}^{+}  \to  n \rho^{+} $&$ 0 $&$ 0 $&$ 0 $&$ 0 $&$ - 2 a_{2} $\\
$ \Xi_{c}^{+}  \to  \Sigma^{0} K^{*+} $&$ 1.268 $&$ 0.296 $&$ -2.491 $&$ 2.232 $&$ - \sqrt{2} a_{1} $\\
$ \Xi_{c}^{+}  \to  \Sigma^{+} K^{*0} $&$ -1.781 $&$ 0.159 $&$ 3.492 $&$ 1.194 $&$ - 2 a_{1} $\\
\hline
$ \Xi_{c}^{0}  \to  \Lambda^{0} K^{*0} $&$ 0.657 $&$ -0.094 $&$ -1.136 $&$ -0.387 $&$ - \frac{\sqrt{6} a_{1}}{3} + \frac{2 \sqrt{6} a_{2}}{3} + \frac{2 \sqrt{6} a_{3}}{3} $\\
$ \Xi_{c}^{0}  \to  p \rho^{-} $&$ 0 $&$ 0 $&$ 0 $&$ 0 $&$ - 2 a_{2} $\\
$ \Xi_{c}^{0}  \to  n \rho^{0} $&$ 0 $&$ 0 $&$ 0 $&$ 0 $&$ \sqrt{2} a_{2} $\\
$ \Xi_{c}^{0}  \to  n \omega $&$ 0 $&$ 0 $&$ 0 $&$ 0 $&$ - 2 \sqrt{2} \tilde{a} + \frac{2 \sqrt{2} a_{1}}{3} - \frac{\sqrt{2} a_{2}}{3} - \frac{2 \sqrt{2} a_{3}}{3} $\\
$ \Xi_{c}^{0}  \to  n \phi $&$ 0 $&$ 0 $&$ 0 $&$ 0 $&$ - 2 \tilde{a} - \frac{4 a_{1}}{3} + \frac{2 a_{2}}{3} + \frac{4 a_{3}}{3} $\\
$ \Xi_{c}^{0}  \to  \Sigma^{0} K^{*0} $&$ 1.255 $&$ -0.114 $&$ -2.454 $&$ -0.839 $&$ \sqrt{2} a_{1} $\\
$ \Xi_{c}^{0}  \to  \Sigma^{-} K^{*+} $&$ 1.788 $&$ 0.425 $&$ -3.502 $&$ 3.138 $&$ - 2 a_{1} $\\
			\hline
			
		\end{tabular}
	\end{center}
\end{table}

\section{Experimental branching ratios }

In Refs.~\cite{Exp:absXic0,Exp:absXicp}, the experimental decay widths are given by
\begin{eqnarray}
\Gamma\left(B^{-} \rightarrow \bar{\Lambda}_{c}^{-} \Xi_{c}^{0}\right)&=&(5.81 \pm 1.39) \times 10^{8}s^{-1}\,,\label{Bm}\\
\Gamma\left(\bar{B}^{0} \rightarrow \bar{\Lambda}_{c}^{-} \Xi_{c}^{+}\right)&=& (7.64 \pm 2.94) \times 10^{8}s^{-1}\,.\label{B0}
\end{eqnarray}
The effective Hamiltonian responsible for the processes is~\cite{Buras:1998raa}
\begin{eqnarray}
{\cal H}_{eff}(\Delta B = 1) = &&\frac{G_F}{\sqrt{2}}\{ \xi_c[ C_1(\mu)Q^c_1(\mu)+C_2(\mu)Q^c_2(\mu)  ] + \xi_u [C_2(\mu)Q^u_2(\mu)   + C_2(\mu)Q^u_u(\mu)]       \nonumber\\
&&-\xi_t \sum_{6}^{i=3}C_i(\mu) Q_i(\mu)
\}\,,
\end{eqnarray}
where $\xi_i \equiv V^*_{ib} V_{is}$, $C_i$ are the Wilson coefficients, and $O_i$ are given as
\begin{eqnarray}
Q_{1}^{q}&=&\left(\bar{b}_{i} q_{j}\right)_{\mathrm{V}-\mathrm{A}}\left(\bar{q}_{j} s_{i}\right)_{\mathrm{V}-\mathrm{A}}\,,\\
Q_{2}^{q}&=&(\bar{b} q)_{\mathrm{V}-\mathrm{A}}(\bar{q} s)_{\mathrm{V}-\mathrm{A}}\,,\\
Q_{3}&=&(\bar{b} s)_{\mathrm{V}-\mathrm{A}} \sum_{q}(\bar{q} q)_{\mathrm{V}-\mathrm{A}}\,,\\
Q_{4}&=&\left(\bar{b}_{i} s_{j}\right)_{\mathrm{V}-\mathrm{A}} \sum_{q}\left(\bar{q}_{j} q_{i}\right)_{\mathrm{V}-\mathrm{A}}\,,\\
Q_{5}&=&(\bar{b} s)_{\mathrm{V}-\mathrm{A}} \sum_{q}(\bar{q} q)_{\mathrm{V}+\mathrm{A}}\,,\\
Q_{6}&=&\left(\bar{b}_{i} s_{j}\right)_{\mathrm{V}-\mathrm{A}} \sum_{q}\left(\bar{q}_{j} q_{i}\right)_{\mathrm{V}+\mathrm{A}}\,,
\end{eqnarray}
with $q = u, d, s,c,b $ in the summations.

The tree order operators, $O_1^c$ and $O_2^c$, are clearly isospin singlet since they do not contain either up or down quark. The penguin operators, $O_3 \sim O_6$, are also isospin singlet since they treat $u$ and $d$ on equal footing. By neglecting $O_1^u$ and $O_2^u$ due to $\xi_u <0.001$, we find that the effective Hamiltonian is an isospin scalar. By using the identity
\begin{equation}
\langle \frac{1}{2} , \frac{1}{2} |_B\left(  |0,0\rangle \otimes | \frac{1}{2},\frac{1}{2}   \rangle\right)_{{\bf B}_c}  =\langle \frac{1}{2} , -\frac{1}{2} |_B\left(  |0,0\rangle \otimes | \frac{1}{2},-\frac{1}{2}\rangle  \right)_{{\bf B}_c}   
\end{equation}
with
\begin{eqnarray}
&&|\frac{1}{2}, \frac{1}{2} \rangle_B  = - |\bar{B} ^0 \rangle \,,\,\,\,\,\,\,|\frac{1}{2}, -\frac{1}{2} \rangle_B  = |B ^- \rangle\,,\\
&&|\frac{1}{2}, \frac{1}{2} \rangle_{{\bf B}_c}  = |\Xi_c^+\rangle \,,\,\,\,\,\,\,\,
|\frac{1}{2}, -\frac{1}{2} \rangle_{{\bf B}_c} = |\Xi_c^0\rangle \,,\,\,\,\,\,\,\, |0,0\rangle_{{\bf B}_c}=|\Lambda^+_c \rangle_{{\bf B}_c}\,,
\end{eqnarray}
we find that the two processes have the same decay widths as stated in Ref.~\cite{Exp:absXicp}.

We average the decay widths in Eqs.~(\ref{B0}) and (\ref{Bm}), given by
\begin{equation}
\Gamma( B \to \Lambda_c^-  \Xi_c ) = (6.14\pm 1.26)\times 10^{8} s^{-1}\,,
\end{equation}
which has a small uncertainty. With $\mathcal{B}\left(\bar{B}^{0} \rightarrow \bar{\Lambda}_{c}^{-} \Xi_{c}^{+}\right) \mathcal{B}\left(\Xi_{c}^{+} \rightarrow \Xi^{-} \pi^{+} \pi^{+}\right) = (3.32 \pm 0.81)\times 10^{-5}$ in Ref.~\cite{Exp:absXicp}, we get
\begin{equation}\label{AbXip}
{\cal B}( \Xi_c^+ \to \Xi^- \pi^+ \pi^+) = (3.56\pm 1.13) \%\,.
\end{equation}

From Eq.~(\ref{AbXip}) and the CLEO data~\cite{XicXirhop}, we have
\begin{equation}
{\cal B}_{ex} ( \Xi_c^+ \to \Xi^0 \pi^+ \pi^0) = ( 8.2 \pm 3.6       )\%\,,
\end{equation}
which contains both resonant and non-resonant contributions.
On the other hand, the latest $SU(3)_F$ analysis with the non-resonance shows that~\cite{Latest_three}
\begin{equation}
{\cal B}_{non}( \Xi_c^+ \to \Xi^0 \pi^+ \pi^0) = ( 1.5 \pm 0.3       )\%\,.
\end{equation}
As a result, the experimental branching ratio is clearly dominated by the resonance. There are two dominating candidates in the resonances, $\Xi_c^+\to \Xi^{0} (1530) \pi^+\,, \Xi^0 \rho^+$. However, the first process is forbidden by the the color symmetry~\cite{Korner,Pati:1970fg,SU(3)Dec},
 which is supported by the experimental data~\cite{XicXirhop}
\begin{equation}
{\cal B} (\Xi_c^+\to \Xi^{0} (1530) \pi^+)/{\cal B} ( \Xi_c^+ \to \Xi^0 \pi^+ \pi^0) < 0.3\,.
\end{equation}
Consequently, we could safely treat the experimental value of  ${\cal B} ( \Xi_c^+ \to \Xi^0 \pi^+ \pi^0)$ 
as ${\cal B} ( \Xi_c^+ \to \Xi^0 \rho^+ )$.

\section*{ACKNOWLEDGMENTS}
We thank Professor Hai-Bo Yu for useful discussions.
This work was supported in part by National Center for Theoretical Sciences and
MoST  (MoST-107-2119-M-007-013-MY3).

\end{document}